\documentclass{article}
\usepackage{setspace}
\usepackage[hidelinks]{hyperref}
\usepackage{amsmath}
\usepackage{parskip} 
\usepackage{natbib}
\usepackage{mathrsfs}
\usepackage{amsthm}

\newtheorem{theorem}{Theorem}[section]
\usepackage{bm}
\usepackage{booktabs}
\usepackage{graphics}
\usepackage{multicol}
\usepackage[bf]{caption}
\usepackage{color}
\usepackage{amssymb}
\usepackage{multirow}
\usepackage{booktabs}
\usepackage{subcaption}

\usepackage{comment}
\usepackage{amsthm}
\usepackage[margin=1.5in]{geometry}
\usepackage{floatrow}
\newfloatcommand{capbtabbox}{table}[][\FBwidth]
\usepackage[titletoc]{appendix}
\usepackage[graphicx]{realboxes}
\usepackage{gensymb}
\usepackage{color}

\usepackage{mathtools}
\title{On relationship between trigonal and cubic symmetry classes of an elasticity tensor
}
\author{Filip P. Adamus\footnote{
Department of Earth Sciences, Memorial University of Newfoundland, Canada, {\tt adamusfp@gmail.com}}}

\date{}
\begin{document}
\maketitle
%%%%%%%%%%%%%%%%%%%%%%
\begin{abstract}
In the literature, there is an ambiguity in defining the relationship between trigonal and cubic symmetry classes of an elasticity tensor. 
We discuss the issue by examining the eigensystems and symmetry groups of trigonal and cubic tensors.
Additionally, we present numerical examples indicating that the sole verification of the eigenvalues can lead to confusion in the identification of the elastic symmetry.
\end{abstract}
%%%%%%%%%%%%%%%%%%%%%%
\section{Introduction}
%%%%%%%%%%%%%%%%%%%%%%
There are eight symmetry classes of an elasticity tensor.
It has been proved, following different approaches, by~\citet{Vianello},~\citet{Chadwick},~\citet{Ting}, or~\citet{BBS2004}.
More than thirty years ago,~\citet{CowinM} discussed the relationship between the elastic symmetries.
They conjectured relations based on numerical experiments.
By adding a symmetry plane to a trigonal material, the authors showed that it reduces to cubic symmetry. 
A more rigorous, but similar approach was followed by~\citet{Chadwick} and~\citet{Ting}.
The above authors slightly adjusted the hierarchy of symmetries proposed by~\citet{CowinM}.
However, they confirmed the statement regarding trigonal and cubic relationship.
Another method, giving analogous conclusions, was shown by~\citet{BBS2007}.
They recognized the relationship between symmetry classes of an elasticity tensor in terms of the eigenvalues and eigenspaces of the associated second-rank tensors.
More recently, similar diagrams of the elastic-symmetry hierarchy were presented by~\citet{Polki},~\citet{Kolev}, and~\citet{Abramian}.

However, in the works of~\citet{BaerheimH},~\citet{BBS2004b},~\citet{BBS2004}, and~\citet{Trusov}, authors do not recognize that trigonal symmetry can be related to cubic.
We want to explain this ambiguity.
We follow the eigensystem approach used by~\citet{BBS2007}.
We find some misprints in their work. 
Also, we treat the problem of the relationship between trigonal and cubic symmetries in a more comprehensive manner.

Additionally, we discuss possible mistakes in the identification of material symmetries.
We present example of tensors exhibiting cubic symmetry that, at first view, look like trigonal.
%%%%%%%%%%%%%%%%%%%%%%
\section{Notation}
%%%%%%%%%%%%%%%%%%%%%%
Elasticity tensor, $c$, is a fourth rank tensor present in Hooke's law,
\begin{equation}\label{eq:one}
\sigma_{ij}=\sum_{k=1}^3\sum_{\ell=1}^3\,c_{ijk\ell}\varepsilon_{kl}\,,\qquad i,\,j\in\{1,\,2,\,3\}\,,
\end{equation}
where $\sigma_{ij}$ is a second-order symmetric stress tensor and $\varepsilon_{kl}$ is a second-order symmetric strain tensor.
The elasticity tensor possesses the following index symmetries,
\begin{gather*}
c_{ijk\ell}=c_{k\ell ij}=c_{ji\ell k}\,.
\end{gather*}
The thirty--six independent components of elasticity tensor can be represented as entries of a $6\times6$ matrix, whereas stress and strain tensors can be viewed as $1\times6$ vectors.
Hence, we can rewrite equation~(\ref{eq:one}) in a  matrix notation~\citep[see, e.g.,][]{SlawinskiRed},
 \begin{equation}\label{eq:two}
\left[
\begin{array}{c}
\sigma_{11}\\
\sigma_{22}\\
\sigma_{33}\\
\sqrt{2}\sigma_{23}\\
\sqrt{2}\sigma_{13}\\
\sqrt{2}\sigma_{12}\\
\end{array}
\right]
=
\left[
\begin{array}{cccccc}
C_{11}&C_{12}& C_{13}& C_{14} & C_{15} & C_{16} \\
C_{12}&C_{22}&C_{23} & C_{24}& C_{25} & C_{26} \\
C_{13}& C_{23} & C_{33} & C_{34} & C_{35} & C_{36} \\
\sqrt{2}C_{14} & \sqrt{2}C_{24} & \sqrt{2}C_{34} & 2C_{44} & 2C_{45} & 2C_{46} \\
\sqrt{2}C_{15} & \sqrt{2}C_{25} & \sqrt{2}C_{35} & 2C_{45} & 2C_{55} & 2C_{56}\\
\sqrt{2}C_{16} & \sqrt{2}C_{26} & \sqrt{2}C_{36} & 2C_{46} & 2C_{56} & 2C_{66}
\end{array}
\right]
\left[
\begin{array}{c}
\varepsilon_{11}\\
\varepsilon_{22}\\
\varepsilon_{33}\\
\sqrt{2}\varepsilon_{23}\\
\sqrt{2}\varepsilon_{13}\\
\sqrt{2}\varepsilon_{12}\\
\end{array}
\right]
\,,
\end{equation}
for which both stress and strain have the same matrix forms.
We denote these six-dimensional vectors as $\hat{\sigma}$ and $\hat{\varepsilon}$.
The relation between the six-dimensional orthonormal basis for matrices from expression~(\ref{eq:two}) and the three-dimensional orthonormal basis for tensors from expression~(\ref{eq:one}) is the following.
\begin{equation}\label{bases}
\hat{e}_\alpha=2^{-\frac{1}{2-\delta_{ij}}}\left(e_i\otimes e_j+e_j\otimes e_i\right)\,,
\end{equation}
where $\alpha=i\delta_{ij}+(1-\delta_{ij})(9-i-j)$ and $\delta_{ij}$ is the Kronecker delta.
The Cartesian base vectors in six dimensions are denoted by $\hat{e}$ and those in
three dimensions by $e$.
To replace pairs $(i,\,j)$ and $(k,\,\ell)$ by single indices in the elasticity tensor from equation~(\ref{eq:two}), we use the same formula as for $\alpha$.

A symmetry of an elasticity tensor, $c$, can be defined as an invariance to the orthogonal transformation of a coordinate system, $A\in O(3)$, where $O(3)$ denotes a group of all orthogonal transformation in three-dimensions.
We denote the set of all symmetries of the elasticity tensor as $G_c$, which is a subgroup of $O(3)$.
Also, a material symmetry, $A\in O(3)$, can be viewed in terms of eigenspaces of the elasticity tensor.
Consider an eigenvalue problem,
\begin{equation}\label{eigen}
\sum_{k=1}^3 \sum_{\ell=1}^3 c_{ijk\ell}\varepsilon_{kl}=\lambda \varepsilon_{ij}\,,
\end{equation}
where ${\bf{\varepsilon}}$ is an eigentensor of $c$ with corresponding eigenvalue $\lambda$.
For a given eigenvalue, we denote the corresponding space of eigentensors by $\Sigma_{\lambda}$. 
$A\in O(3)$ is a symmetry of $c$ if and only if it preserves $\Sigma_{\lambda}$ for all eigenvalues of $c$, as discussed by~\citet{BBS2007}.
 
Finally, it is important to define the symmetry class.
Two elasticity tensors $c_1$ and $c_2$ belong to the same symmetry class if their symmetry groups are orthogonally conjugate.
The orthogonal conjugacy means that there exists a transformation, $A\in O(3)$, such that $G_{c_2}=AG_{c_1} A^T$, where $T$ denotes transposition.
%%%%%%%%%%%%%%%%%%%%%%
\section{Relationship between trigonal and cubic symmetries}
%%%%%%%%%%%%%%%%%%%%%%
%%%%%%%%%%%%%%%%%%%%%%
\subsection{Particular case}
%%%%%%%%%%%%%%%%%%%%%%
Let us consider an elasticity tensor that has trigonal symmetry, meaning that is invariant under three-fold rotation of the coordinate system.
Let us assume that base vector $e_3$ is parallel to the rotations.
The matrix representation of this tensor, with respect to the natural coordinate system, is
\begin{equation}\label{matrix:three}
C^{\rm {t}}
=
\left[
\begin{array}{cccccc}
C_{11}&C_{12} & C_{13} &\sqrt{2}C_{14} & 0  & 0  \\
C_{12}&C_{11} &C_{13}& -\sqrt{2}C_{14}& 0  & 0  \\
C_{13}&C_{13}& C_{33} & 0 & 0 & 0  \\
\sqrt{2}C_{14} & -\sqrt{2}C_{14}& 0 &2C_{44}& 0 & 0 \\
0& 0  & 0 & 0 & 2C_{44}& 2C_{14}\\
0 & 0  & 0  & 0 & 2C_{14} &C_{11}-C_{12}
\end{array}
\right].
\end{equation}
\begin{comment}
that can be obtained from $\tan(3\theta_1)=C_{14}/
C_{15}$ and $\tan(3\theta_2)=C_{15}/C_{14}$, then matrix~(\ref{matrix:three}) reduces the number of independent parameters to six, since $C_{15}=0$ or $C_{14}=0$, respectively.
In such a case, the trigonal tensor is expressed with respect to the natural basis.
\end{comment}
Herein, we focus on the above particular case of a trigonal symmetry that has six independent parameters.
The eigenvalues of matrix~(\ref{matrix:three}) are
\begin{align*}
\lambda_1&=\frac{1}{2}\left[C_{11}+C_{12}+C_{33}+\sqrt{\left(C_{11}+C_{12}-C_{33}\right)^2+8C_{13}^2}\,\right]\,,\\
\lambda_2&=\frac{1}{2}\left[C_{11}+C_{12}+C_{33}-\sqrt{\left(C_{11}+C_{12}-C_{33}\right)^2+8C_{13}^2}\,\right]\,,\\
\lambda_3&=\frac{1}{2}\left[C_{11}-C_{12}+2C_{44}+\sqrt{\left(C_{11}-C_{12}-2C_{44}\right)^2+16C_{14}^2}\,\right]\,,\\
\lambda_4&=\frac{1}{2}\left[C_{11}-C_{12}+2C_{44}-\sqrt{\left(C_{11}-C_{12}-2C_{44}\right)^2+16C_{14}^2}\,\right]\,,
\end{align*}
where the multiplicities of $\lambda$'s are $m_1=1$, $m_2=1$, $m_3=2$, and $m_4=2$, respectively.
Below, we use the auxiliary parameters discussed by~\citet{BBS2007}, namely,
\begin{align}\label{gam1}
\gamma_1&:=-\frac{C_{11}+C_{12}-C_{33}+\sqrt{(C_{11}+C_{12}-C_{33})^2+8C_{13}^2}}{2C_{13}}\,,\\\label{gam2}
\gamma_2&:=\frac{C_{11}-C_{12}-2C_{44}+\sqrt{(C_{11}-C_{12}-2C_{44})^2+16C_{14}^2}}{4C_{14}}\,.
\end{align}

A tensor with cubic symmetry is invariant under two rotations by $\pi/2$ around two mutually orthogonal axes.
In this section, we want to examine if upon introducing certain dependencies among entries of $C^{\rm {t}}$, the elasticity tensor can specialize to cubic symmetry. 
To do so, we invoke the coordinate-free conditions to identify the cubic symmetry of an elasticity tensor from the work of~\citet{BBS2007}.

\hphantom{x}
\begin{theorem}
Consider an elasticity tensor, $c$, for which the following conditions are satisfied.
\begin{enumerate}
\item{$c$ has three distinct eigenvalues, $\lambda_1$, $\lambda_2$ and $\lambda_3$, with corresponding multiplicities, $\,\,\,\,m_1=1$, $m_2=2$ and $m_3=3$,}
\item{the corresponding spaces of eigentensors $\Sigma_1$ and $\Sigma_2$ are such that}
\begin{enumerate}
\item{all $\varepsilon$ in $\Sigma_1$ have eigenvalues with multiplicity three.}
\item{all $\varepsilon$ in $\Sigma_2$ have three common eigenvectors.}
\end{enumerate}
\end{enumerate}
Such an elasticity tensor has cubic symmetry. 
Also, the three common eigenvectors of $\varepsilon$ in $\Sigma_2$ determine a natural basis of $c$.
\end{theorem}

Is it possible to find a particular case where $C^{\rm {t}}$ satisfies all the points of Theorem 3.1?
In other words, can a nominally trigonal tensor have cubic symmetry?
Let us check it.

To make $C^{\rm {t}}$ satisfy the first point of the theorem, either $\lambda_1=\lambda_3$, or $\lambda_2=\lambda_4$.
These conditions are respectively tantamount to
\begin{equation*}
C_{33}-2C_{44}=C_{13}\,\gamma_1+2C_{14}\,\gamma_2\,\qquad \rm{and}\, \qquad
2C_{12}=-C_{13}\,\gamma_1-2C_{14}\,\gamma_2\,,
\end{equation*}
where we have used the auxiliary parameters from expressions~(\ref{gam1}) and (\ref{gam2}).
%Let us assume the first possibility.
To examine the second point of Theorem 3.1, we need to solve the eigenvalue problem from equation~(\ref{eigen}).
To do so, we follow a simple strategy.
First, we consider the problem in six dimensions using the relation among the bases from~(\ref{bases}).
As a consequence, we solve
$(C^{\rm{t}}-I\lambda)\hat{\varepsilon}=0$.
Then, again we use (\ref{bases}) and express the resulting $1\times6$ eigenvectors $\hat{\varepsilon}$, as $3\times3$ eigentensors $\varepsilon$.
In this way, we are able to examine the eigenvalues of $\varepsilon$ and check if point $2$ of of Theorem 3.1 is obeyed.
Hence, we get a system of six equations,
\begin{align}
%\begin{aligned}
(C_{11}-\lambda)\varepsilon_{11}+C_{12}\varepsilon_{22}+C_{13}\varepsilon_{33}+2C_{14}\varepsilon_{23}=0\,,\label{system:1}\\
C_{12}\varepsilon_{11}+(C_{11}-\lambda)\varepsilon_{22}+C_{13}\varepsilon_{33}-2C_{14}\varepsilon_{23}=0\,,\\
C_{13}\varepsilon_{11}+C_{13}\varepsilon_{22}+(C_{33}-\lambda)\varepsilon_{33}=0\,,\label{system:3}\\
\sqrt{2}C_{14}\varepsilon_{11}-\sqrt{2}C_{14}\varepsilon_{22}+(2C_{44}-\lambda)\sqrt{2}\varepsilon_{23}=0\,,\\
(2C_{44}-\lambda)\sqrt{2}\varepsilon_{13}+2\sqrt{2}C_{14}\varepsilon_{12}=0\,,\\
2\sqrt{2}C_{14}\varepsilon_{13}+(C_{11}-C_{12}-\lambda)\sqrt{2}\varepsilon_{12}=0\,.\label{system:6}
%\end{aligned}
\end{align}
First, let us focus on point $2(a)$ of Theorem 3.1. 
The eigenvalues of $3\times3$ symmetric matrix $\varepsilon$ have multiplicity three if and only if matrix is diagonal and its entries equal to each other, which can be easily proven. 
Hence, to satisfy point $2(a)$, we require $\varepsilon_{11}=\varepsilon_{22}=\varepsilon_{33}$ and  $\varepsilon_{23}=\varepsilon_{13}=\varepsilon_{12}=0$.
We assume that ${\varepsilon}$ is not a zero matrix. 
In other words, we do not consider an undeformed state.
Combining equations~(\ref{system:1}) and~(\ref{system:3}), we get relation 
\begin{equation}\label{relation}
C_{13}-C_{12}=C_{11}-C_{33}\,.
\end{equation}
To satisfy Theorem 3.1, $C_{13}$ cannot equal to zero. Note that if $C_{13}=0$, then $\lambda_1=\lambda_2$.
Assuming that $\lambda_1=\lambda_3$ and using relation~(\ref{relation}), we simplify equation~(\ref{system:1}) and get 
\begin{equation*}
\lambda_2=C_{11}+C_{12}+C_{13}\qquad\rightarrow\qquad 3C_{13}+\sqrt{(3C_{13})^2}=0\,.
\end{equation*}
We see that $C_{13}$ must be negative.
Relation~(\ref{relation}) and $C_{13}<0$ is tantamount to $\gamma_1=1$.
Analogously, if $\lambda_2=\lambda_4$, then $C_{13}$ must be positive.
If $C_{13}>0$, then $\gamma_1$ must be equal to negative two.
Now, let us focus on point $2(b)$ of Theorem 3.1. In other words, we consider the case of $\lambda$ with multiplicity two and its corresponding space of eigentensors $\Sigma_2$.
We solve equations~(\ref{system:1})--(\ref{system:6}) and get 
\begin{align*}
\Sigma_2&=
\begin{bmatrix}
\varepsilon_{11} & -\varepsilon_{13}\left(\dfrac{2C_{44}-\lambda}{2C_{14}}\right) & \varepsilon_{13}  \\
-\varepsilon_{13}\left(\dfrac{2C_{44}-\lambda}{2C_{14}}\right)&-\varepsilon_{11}&-\varepsilon_{11}\left(\dfrac{2C_{14}}{2C_{44}-\lambda}\right)  \\
\varepsilon_{13}&-\varepsilon_{11}\left(\dfrac{2C_{14}}{2C_{44}-\lambda}\right)& \varepsilon_{33}(C_{33}-\lambda)  \\
\end{bmatrix}
\\
\\
&=a_1
\begin{bmatrix}
&&\\[-0.3cm]
0 & \dfrac{-2C_{44}+\lambda}{2C_{14}} & 1  \\[0.28cm]
\dfrac{-2C_{44}+\lambda}{2C_{14}}&0&0  \\[0.28cm]
1&0& 0  \\
\end{bmatrix}
+
a_2
\begin{bmatrix}
&&\\[-0.3cm]
1 & 0 & 0  \\[0.2cm]
0&-1&\dfrac{-2C_{14}}{2C_{44}-\lambda}  \\[0.2cm]
0&\dfrac{-2C_{14}}{2C_{44}-\lambda}& 0 
\end{bmatrix}\,,
\end{align*}
where $a_{i}$ are constants and 
\begin{equation*}
\frac{2C_{14}}{2C_{44}-\lambda}=\frac{C_{11}-C_{12}-\lambda}{2C_{14}}\,.
\end{equation*}
In order not to have three independent eigentensors---without loss of generality---we assume that $\varepsilon_{33}=0$.
Also, we notice that $C_{14}\neq0$, since if $C_{14}=0$ then matrix~(\ref{matrix:three}) has transversely-isotropic symmetry. 
As a result, $\lambda_j\neq2C_{44}$ or $\lambda_j\neq C_{11}-C_{12}$, where $j\in(3,4)$.
A pair of $3\times3$ matrices have three common eigenvectors if and only if they are commutative.
To find conditions satisfying point $2(b)$ of Theorem 3.1,
let us use the auxiliary parameter $\gamma_2$, which can be rewritten in terms of $\lambda_3$ or $\lambda_4$,
\begin{equation*}
\gamma_2=\frac{\lambda_3-2C_{44}}{2C_{14}}=\frac{C_{11}-C_{12}-\lambda_4}{2C_{14}}\,.
\end{equation*}
If $\lambda_1=\lambda_3$, then we can write
\begin{equation*}
\Sigma_2=
a_1\left[
\begin{array}{ccc}
0 & -\gamma_2^{-1}& 1  \\
 -\gamma_2^{-1}&0&0  \\
1&0& 0  \\
\end{array}
\right]
+
a_2\left[
\begin{array}{ccc}
1 & 0 & 0  \\
0&-1&-\gamma_2 \\
0&-\gamma_2& 0 \\
\end{array}
\right]
\end{equation*}
and the two matrices are commutative if $\gamma_2=\pm\sqrt{2}$. 
If $\lambda_2=\lambda_4$, then we get
\begin{equation*}
\Sigma_2=
a_1\left[
\begin{array}{ccc}
0 & \gamma_2& 1  \\
 \gamma_2&0&0  \\
1&0& 0  \\
\end{array}
\right]
+
a_2\left[
\begin{array}{ccc}
1 & 0 & 0  \\
0&-1&\gamma_2^{-1} \\
0&\gamma_2^{-1}& 0 
\end{array}
\right]
\end{equation*}
and the two matrices are commutative if $\gamma_2=\pm1/\sqrt{2}$. 
To sum up, matrix~(\ref{matrix:three}) that nominally represents a trigonal symmetry can be specialized to cubic if 
\begin{equation}\label{eq:cond}
\begin{cases}
\lambda_1=\lambda_3\\
\gamma_1=1\\
\gamma_2=\pm\sqrt{2}
\end{cases} 
\qquad \rm{or} \qquad\,
\begin{cases}
\lambda_2=\lambda_4\\
\gamma_1=-2\\
\gamma_2=\pm1/\sqrt{2}
\end{cases} \,,
\end{equation}
which are the necessary and sufficient conditions.
They differ from the ones shown by~\citet{BBS2007} due to the possible misprint present in their work.
Also, we can express conditions~(\ref{eq:cond}) in terms of the relations between elasticity parameters,
\begin{equation}\label{sixteen}
\begin{cases}
C_{33}-2C_{44}=C_{13}\pm2\sqrt{2}C_{14}\\
C_{13}=C_{11}+C_{12}-C_{33}<0\\
2C_{44}
=C_{11}-C_{12}\pm\sqrt{2}C_{14}
\end{cases} 
\qquad \rm{or} \qquad\,
\begin{cases}
2C_{13}=2C_{12}\pm\sqrt{2}C_{14}\\
C_{13}=C_{11}+C_{12}-C_{33}>0\\
\pm\sqrt{2}C_{14}=C_{11}-C_{12}-2C_{44}
\end{cases} \,,
\end{equation}
respectively. 
If $C_{14}>0$ then $\pm$ changes to plus, or if $C_{14}<0$ then $\pm$ changes to minus.
The relations on the right-hand part of expression~(\ref{sixteen}) are in accordance with~\citet{Ting}.

Consider a special case of $C^t$, where $\lambda_1=\lambda_3$, $\gamma_1=1$, $\gamma_2=-\sqrt{2}$.
The natural basis of such a cubic tensor is---according to Theorem 3.1---determined by an orthogonal transformation,
\begingroup
\renewcommand*{\arraystretch}{1.5}
\begin{equation}\label{ahat}
\hat{A}
=
\begin{bmatrix}
\frac{\sqrt{2}}{2}&\frac{\sqrt{6}}{6} & \frac{\sqrt{3}}{3} \\
-\frac{\sqrt{2}}{2}&\frac{\sqrt{6}}{6} &\frac{\sqrt{3}}{3} \\
0&-\frac{\sqrt{6}}{3}& \frac{\sqrt{3}}{3}
\end{bmatrix}
=
\begin{bmatrix}
\frac{\sqrt{2}}{2}&\frac{\sqrt{2}}{2}& 0 \\
-\frac{\sqrt{2}}{2}&\frac{\sqrt{2}}{2} &0 \\
0&0& 1
\end{bmatrix}
\begin{bmatrix}
1&0 & 0 \\
0&\frac{\sqrt{3}}{3} &\frac{\sqrt{6}}{3} \\
0&-\frac{\sqrt{6}}{3}& \frac{\sqrt{3}}{3}
\end{bmatrix}
=R_{\pi/4,\,e_3}\,R_{\arccos\left(\frac{\sqrt{3}}{3}\right), \,e_1}
\end{equation}
\endgroup
that is composed from three common eigenvectors in $\Sigma_2$.
Transformation $\hat{A}$ can be understood, for instance, as a rotation about $e_1$ by an angle $\theta=\arccos(\frac{\sqrt{3}}{3})$ of the coordinate system that was previously rotated by $\theta=\pi/4$ about $e_3$.
$R_{\theta,\,e_i}$ denotes rotation around $e_i$.
%%%%%%%%%%%%%%%%%%%%%%
\subsection{General case}
%%%%%%%%%%%%%%%%%%%%%%
Let us consider a cubic tensor expressed with respect to the natural basis. 
The base vectors $e_1$, $e_2$ of such a tensor are parallel to two orthogonal coordinate axes.
 Its symmetry group is
 %
 %\begin{align*}
 %G_c=\{A\in O(3)\,,\,\, Ae_i=\pm e_j\,;\,\, i,\,j\,\in\{1,2,3\}\,\} \,.
 %\end{align*}
 %
\begin{align*}
 G_c&=\{\, \pm I,\,\pm R_{\pm\pi/2,\,e_i},\,\pm M_{e_i},\,\pm M_{\left[1,\,1,\,0\right]},\,\pm M_{\left[1,\,-1,\,0\right]}\,
 \\
 &\qquad\pm M_{\left[1,\,0,\,1\right]},
 \,\pm M_{\left[1,\,0,\,-1\right]},
\,\pm M_{\left[0,\,1,\,1\right]}
\,\pm M_{\left[0,\,1,\,-1\right]}\,; \quad i\,\in\{1,2,3\}\,\}\,,
 \end{align*}
 where $M$ denotes a reflection about a plane with a normal vector indicated by the subscript.
From the previous section, we infer that $G_c$ is orthogonally conjugate to the symmetry group of $C^t$, which we denote as $\hat{G}_c$.
In other words,
\begin{equation*}
G_c=\hat{A}\,\hat{G}_c\,\hat{A}^T\,,
\end{equation*}
which we can rewrite as
\begin{equation*}
\hat{A}^T\,G_c\,\hat{A}=\hat{G}_c\,,
\end{equation*}
where $\hat{A}$ stands for transformation~(\ref{ahat}).
Let us find some elements of the group $\hat{G}_c$.
For instance,
\begin{align}\label{first}
&\hat{A}^T\,(\pm I)\,\hat{A}=\pm I\,,\\ 
&\hat{A}^T\,(\pm I)\,M_{[1,\,-1,\,0]}\,\hat{A}=\pm M_{e_1}\,,\\
&\hat{A}^T\,(\pm I)\,M_{[1,\,0,\,-1]}\,\hat{A}=\pm M_{[\cos(\pi/3),\,\sin(\pi/3),\,0]}\,,\\
&\hat{A}^T\,(\pm I)\,M_{[0,\,1,\,-1]}\,\hat{A}=\pm M_{[\cos(2\pi/3),\,\sin(2\pi/3),\,0]}\,,\\
&\hat{A}^T\,(\pm I)\,M_{[1,\,0,\,1]}\,R_{\pi/2,\,e_3}\hat{A}=\mp R_{2\pi/3,\,e_3}\,,\\
&\hat{A}^T\,(\pm I)\,M_{[1,\,0,\,1]}\,R_{-\pi/2,\,e_3}\hat{A}=\mp R_{-2\pi/3,\,e_3}\,,\label{prelast} \\ 
&\hat{A}^T\,M_{[1,\,1,\,0]}\,\hat{A}=M_{[0,\,\sqrt{3}/3,\,\sqrt{6}/3]}\,, \label{last}
\end{align}
and so on.
Symmetry group of a trigonal tensor expressed in a natural coordinate system is
 \begin{equation*}
 G_t=\{\, \pm I,\,\pm R_{\pm2\pi/3,\,e_3},\,\pm M_{e_1},\,\pm M_{\left[\cos(\pi/3),\,\sin(\pi/3),\,0\right]},\,\pm M_{\left[\cos(2\pi/3),\,\sin(2\pi/3),\,0\right]}\,\}\,.
 \end{equation*}
We easily notice that $G_t$ is not a subgroup of $G_c$. 
However, $G_t\subset \hat{G}_c$; note that the transformations~(\ref{first})--(\ref{prelast}) form $G_t$, but reflection~(\ref{last}) is included in $\hat{G}_c$, not in $G_t$.
Thus, symmetry group of a trigonal tensor expressed with respect to natural basis is a subgroup of an orthogonally-conjugate cubic symmetry group.
The relation between groups is true for any orientation of the coordinate system, since $A\,G_t\,A^T\subset A\,\hat{G}_c\,A^T$, where $A\in O(3)$.
We can state that---in view of the orthogonal conjugation--cubic symmetry class contains trigonal symmetry class.
%%%%%%%%%%%%%%%%%%%%%%
\section{Numerical examples}
%%%%%%%%%%%%%%%%%%%%%%
Consider matrix~(\ref{matrix:three}).
At first view, it looks as if it represents a trigonal symmetry.
However, as we have shown above, instead, it may represent a cubic symmetry.
A trigonal symmetry group is also a subgroup of transversely-isotropic and isotropic symmetries.
Hence, matrix~(\ref{matrix:three}) may represent one out of four possible symmetry classes.
Let us examine two numerical examples of such a matrix and check its symmetry.

Consider first example,
\begin{equation*}
C_1
=
\left[
\begin{array}{cccccc}
10-\frac{\sqrt{2}}{2}&\frac{\sqrt{2}}{2}-1& -1&-\sqrt{2} & 0 & 0  \\
\frac{\sqrt{2}}{2}-1&10-\frac{\sqrt{2}}{2}&-1& \sqrt{2}& 0  & 0  \\
-1&-1& 10 & 0 & 0 & 0  \\
-\sqrt{2}& \sqrt{2} & 0 &11-2\sqrt{2}& 0 & 0  \\
0& 0  & 0 & 0 & 11-2\sqrt{2}& -2\\
0 & 0  & 0  & 0 & -2 & 11-\sqrt{2}
\end{array}
\right].
\end{equation*}
Such a matrix has the following eigenvalues
\begin{equation*}
\lambda_2=8\,,\quad \lambda_4=6.7574\,,\quad {\rm{and}}\quad \lambda_1= \lambda_3=11\,
\end{equation*}
with multiplicities $m_1=1$, $m_2=2$, and $m_3=3$, respectively.
The multiplication and number of distinct eigenvalues, according to Theorem 3.1, corresponds to cubic symmetry.
To make sure that $C_1$ represents a tensor that has cubic symmetry, we check the eigentensor spaces $\Sigma_1$ and $\Sigma_2$ along with corresponding eigensystems:
%\begin{equation*}
\begin{align*}
 \Sigma_1&=a_0
\begin{bmatrix}
&&\\[-0.28cm]
-\frac{\sqrt{3}}{3} & 0 & 0 \\
0&-\frac{\sqrt{3}}{3} &0  \\
0&0& -\frac{\sqrt{3}}{3}   \\[0.11cm]
\end{bmatrix}
\qquad\qquad\xrightarrow{\text{corresponding eigensystem}} \quad
&\begin{matrix}
 -\frac{\sqrt{3}}{3} \,\, [1,0,0]\,\,\\[0.15cm]
  -\frac{\sqrt{3}}{3} \,\, [0,1,0]\,,\\[0.15cm]
  -\frac{\sqrt{3}}{3} \,\, [0,0,1]\,\,\\[0.4cm]
 \end{matrix}
 \\
\Sigma_2&=a_1
\begin{bmatrix}
&&\\[-0.25cm]
-\frac{\sqrt{6}}{6}  & 0& 0 \\
0&\frac{\sqrt{6}}{6} &-\frac{\sqrt{3}}{3} \\[0.1cm]
0&-\frac{\sqrt{3}}{3} & 0  \\[0.11cm]
\end{bmatrix}
\qquad\qquad\xrightarrow{\text{corresponding eigensystem}} \quad
&\begin{matrix}
-\frac{\sqrt{6}}{6} \,\, \left[\frac{\sqrt{2}}{2},\frac{\sqrt{6}}{6},\frac{\sqrt{3}}{3}\right]\\[0.2cm]
-\frac{\sqrt{6}}{6}\,\, \left[-\frac{\sqrt{2}}{2}, \frac{\sqrt{6}}{6},\frac{\sqrt{3}}{3}\right]\\[0.2cm]
\frac{\sqrt{6}}{3}\,\, \left[0,-\frac{\sqrt{6}}{3},\frac{\sqrt{3}}{3}\right]\\[0.2cm]
\end{matrix}
\\
&+a_2
\begin{bmatrix}
&&\\[-0.25cm]
0 & -\frac{\sqrt{6}}{6}  & -\frac{\sqrt{3}}{3}   \\
-\frac{\sqrt{6}}{6} &0&0  \\[0.11cm]
-\frac{\sqrt{3}}{3}  &0& 0 \\[0.11cm]
\end{bmatrix}
\qquad\qquad\xrightarrow{\text{corresponding eigensystem}} \quad
&\begin{matrix}
-\frac{\sqrt{6}}{2} \,\, \left[\frac{\sqrt{2}}{2},\frac{\sqrt{6}}{6},\frac{\sqrt{3}}{3}\right]\\[0.2cm]
\frac{\sqrt{6}}{2}\,\, \left[-\frac{\sqrt{2}}{2}, \frac{\sqrt{6}}{6},\frac{\sqrt{3}}{3}\right]\\[0.2cm]
\,\,\,0\,\,\, \left[0,-\frac{\sqrt{6}}{3},\frac{\sqrt{3}}{3}\right]\\
\end{matrix}\,.
\end{align*}
%\end{equation*}
All conditions of Theorem 3.1 are satisfied. 
$C_1$ represents a tensor with cubic symmetry.
Note that eigenvectors $\left[\frac{\sqrt{2}}{2},\frac{\sqrt{6}}{6},\frac{\sqrt{3}}{3}\right]$, $\left[-\frac{\sqrt{2}}{2}, \frac{\sqrt{6}}{6},\frac{\sqrt{3}}{3}\right]$, and $\left[0,-\frac{\sqrt{6}}{3},\frac{\sqrt{3}}{3}\right]$ form transformation matrix~(\ref{ahat}).
Also, conditions~(\ref{eq:cond}) are obeyed, as expected.

Consider a second example,
\begin{equation*}
C_2
=
\begin{bmatrix}
&&&&&\\[-0.3cm]
\frac{23-\sqrt{2}}{2}&\frac{\sqrt{2}-1}{2}& 1&-\sqrt{2} & 0 & 0  \\[0.12cm]
\frac{\sqrt{2}-1}{2}&\frac{23-\sqrt{2}}{2}&1& \sqrt{2}& 0  & 0  \\[0.12cm]
1&1& 10 & 0 & 0 & 0  \\[0.12cm]
-\sqrt{2}& \sqrt{2} & 0 &12-2\sqrt{2}& 0 & 0  \\[0.12cm]
0& 0  & 0 & 0 & 12-2\sqrt{2}& -2\\[0.12cm]
0 & 0  & 0  & 0 & -2 & 12-\sqrt{2}\\[0.12cm]
\end{bmatrix}.
\end{equation*}
Such a matrix has the following eigenvalues
\begin{equation*}
\lambda_2=9\,,\quad \lambda_4=7.7574\,,\quad {\rm{and}}\quad \lambda_1= \lambda_3=12\,
\end{equation*}
with multiplicities $m_1=1$, $m_2=2$, and $m_3=3$, respectively.
Again, the eigenvalues suggest that the tensor has cubic symmetry.
Hovewer, if we check the eigentensor spaces and  corresponding eigensystems,
\begin{align*}
 \Sigma_1&=a_0
\begin{bmatrix}
&&\\[-0.28cm]
-\frac{\sqrt{6}}{6} & 0 & 0 \\
0&-\frac{\sqrt{6}}{6} &0  \\
0&0& \frac{\sqrt{6}}{3}   \\[0.11cm]
\end{bmatrix}
\qquad\qquad\xrightarrow{\text{corresponding eigensystem}} \,\,\,
&\begin{matrix}
 -\frac{\sqrt{6}}{6} \,\, [1,0,0]\,\,\\[0.15cm]
  -\frac{\sqrt{6}}{6} \,\, [0,1,0]\,,\\[0.15cm]
  \,\,\frac{\sqrt{6}}{3} \,\, [0,0,1]\,\,\\[0.4cm]
 \end{matrix}
 \\
\Sigma_2&=a_1
\begin{bmatrix}
&&\\[-0.25cm]
-\frac{\sqrt{6}}{6}  & 0& 0 \\
0&\frac{\sqrt{6}}{6} &-\frac{\sqrt{3}}{3} \\[0.1cm]
0&-\frac{\sqrt{3}}{3} & 0  \\[0.11cm]
\end{bmatrix}
\qquad\qquad\xrightarrow{\text{corresponding eigensystem}} \quad
&\begin{matrix}
-\frac{\sqrt{6}}{6} \,\, \left[\frac{\sqrt{2}}{2},\frac{\sqrt{6}}{6},\frac{\sqrt{3}}{3}\right]\\[0.2cm]
-\frac{\sqrt{6}}{6}\,\, \left[-\frac{\sqrt{2}}{2}, \frac{\sqrt{6}}{6},\frac{\sqrt{3}}{3}\right]\\[0.2cm]
\frac{\sqrt{6}}{3}\,\, \left[0,-\frac{\sqrt{6}}{3},\frac{\sqrt{3}}{3}\right]\\[0.2cm]
\end{matrix}
\\
&+a_2
\begin{bmatrix}
&&\\[-0.25cm]
0 & -\frac{\sqrt{6}}{6}  & -\frac{\sqrt{3}}{3}   \\
-\frac{\sqrt{6}}{6} &0&0  \\[0.11cm]
-\frac{\sqrt{3}}{3}  &0& 0 \\[0.11cm]
\end{bmatrix}
\qquad\qquad\xrightarrow{\text{corresponding eigensystem}} \quad
&\begin{matrix}
-\frac{\sqrt{6}}{2} \,\, \left[\frac{\sqrt{2}}{2},\frac{\sqrt{6}}{6},\frac{\sqrt{3}}{3}\right]\\[0.2cm]
\frac{\sqrt{6}}{2}\,\, \left[-\frac{\sqrt{2}}{2}, \frac{\sqrt{6}}{6},\frac{\sqrt{3}}{3}\right]\\[0.2cm]
\,\,\,0\,\,\, \left[0,-\frac{\sqrt{6}}{3},\frac{\sqrt{3}}{3}\right]\\
\end{matrix}\,,
\end{align*}
we notice that point $2(a)$ of Theorem 3.1 is not satisfied. Thus, matrix $C_2$ does not represent a cubic tensor.
As expected, conditions~(\ref{eq:cond}) are not obeyed, since $\lambda_1=\lambda_3$, $\gamma_1=-2$, and $\gamma_2=-\sqrt{2}$.
Using other theorems from~\citet{BBS2007}, we infer that $C_2$ represents trigonal symmetry.
%%%%%%%%%%%%%%%%%%%%%%
\section{Conclusions}
%%%%%%%%%%%%%%%%%%%%%%
First, we use a particular example of a matrix that nominally represents a trigonal tensor. 
Based on a theorem from~\citet{BBS2007}, we show the conditions to make the aforementioned matrix represent a cubic tensor.
These conditions differ from the ones shown in~\citet{BBS2007}, but some of them are in accordance with~\citet{Ting}.

\begin{figure}[h]
\centering
\includegraphics[width=8.2cm]{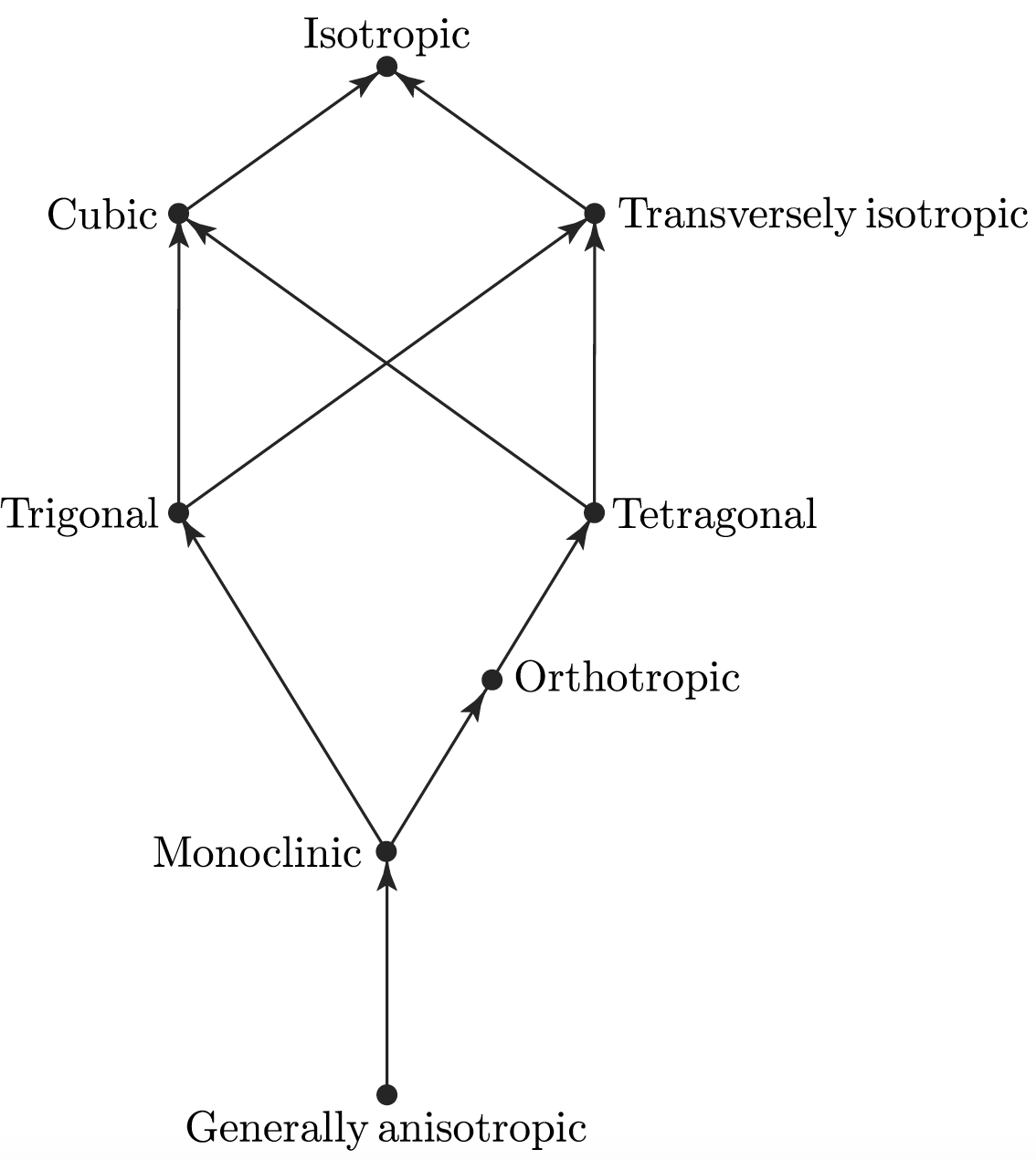}
\caption{\footnotesize{Relations among symmetry classes of an elasticity tensor.}}
\label{fig1}
\end{figure}

Further, from the particular example, we proceed to a general case. 
We show that cubic symmetry class contains trigonal symmetry class.
This is the consequence of the fact that a trigonal symmetry group is a subgroup of the orthogonally-conjugate cubic symmetry group.
Certain authors do not notice the relationship between trigonal and cubic symmetries.
It can be caused by neglecting the orthogonal conjugacy among symmetry groups.
Correct relations among symmetry classes of an elasticity tensor are shown in Figure~\ref{fig1}.
%Therein, we indicate a possible way of imposing conditions on elasticity parameters that result in desired relations among eight classes.   

Lastly, we present two numerical examples of matrices that---at first view---pretend to represent a tensor having trigonal symmetry class.
Following the theorem from~\citet{BBS2007}, we show that to correctly recognize the symmetry class of a tensor, it is crucial to examine not only its eigenvalues, but also eigentensors.
Forgetting about the eigentensors may lead to misidentification of a tensor symmetry. 

%%%%%%%%%%%%%%%%%%%%%%
\section*{Acknowledgements}
%%%%%%%%%%%%%%%%%%%%%%
We wish to acknowledge discussions with Andrej B\'ona and Michael A. Slawinski.
Also, we thank David Dalton for proofreading the article.
The research was done in the context of The Geomechanics Project supported by Husky Energy.
%%%%%%%%%%%%%%%%%%%%%%
%%%%%%%%%%%%%%%%%%%%%%
\bibliography{trig}
\bibliographystyle{apa}
%%%%%%%%%%%%%%%%%%%%%%v
\newpage
\begin{comment}
RELATIONS BETWEEN PARAMETERS TO BE PLACED NEXT OR OVER ARROWS IN FIG. 1. They can be written with comas, for convenience.

gen--mon
\begin{equation*}
\begin{aligned}
C_{15}=C_{25}=C_{35}=C_{45}=0 \\
C_{16}=C_{26}=C_{36}=C_{46}=0 \\
\end{aligned}
\end{equation*}
mon--trig
\begin{equation*}
\begin{aligned}
&C_{34}=0 \\
&C_{11}=C_{22} \\
&C_{13}=C_{23} \\
&C_{44}=C_{55} \\
&C_{14}=-C_{24}=C_{56} \\
&C_{11}-C_{12}=2C_{66}\\
\end{aligned}
\end{equation*}
trig--cubic
\begin{equation*}
\begin{aligned}
2C_{13}=2C_{12}+\sqrt{2}C_{14}\\
C_{13}=C_{11}+C_{12}-C_{33}>0\\
\sqrt{2}C_{14}=C_{11}-C_{12}-2C_{44}
\end{aligned}
\end{equation*}
trig--TI (transversely-isotropic)
\begin{equation*}
\begin{aligned}
C_{14}=0
\end{aligned}
\end{equation*}
cubic--isotropic
\begin{equation*}
\begin{aligned}
C_{11}-C_{12}=2C_{44}
\end{aligned}
\end{equation*}

mon--ort
\begin{equation*}
\begin{aligned}
C_{14}=C_{24}=C_{34}=C_{56}=0
\end{aligned}
\end{equation*}
ort--tetragonal
\begin{equation*}
\begin{aligned}
C_{11}=C_{22}\\
C_{13}=C_{23}\\
C_{44}=C_{55}\\
\end{aligned}
\end{equation*}
tetragonal--cubic
\begin{equation*}
\begin{aligned}
C_{11}=C_{33}\\
C_{12}=C_{13}\\
C_{44}=C_{66}\\
\end{aligned}
\end{equation*}
tetragonal--TI
\begin{equation*}
\begin{aligned}
C_{11}-C_{12}=2C_{66}
\end{aligned}
\end{equation*}
TI--isotropic
\begin{equation*}
\begin{aligned}
&C_{11}=C_{33}\\
&C_{12}=C_{13}\\
&C_{11}-C_{12}=2C_{44}
\end{aligned}
\end{equation*}
\end{comment}
%%%%%%%%%%%%%%%%%%%%%%
%%%%%%%%%%%%%%%%%%%%%%v
\end{document}